\documentclass[aps,prb,superscriptaddress,floatfix,10pt,twocolumn]{revtex4-2}
\usepackage{graphicx,graphics}
\usepackage{dcolumn}
\usepackage{amsmath,amssymb,amsfonts}
\usepackage{latexsym,verbatim}
\usepackage{bm}
\usepackage{color}
\usepackage{ulem}
\usepackage{multirow}
\usepackage[abs]{overpic}
\usepackage[breaklinks=true,colorlinks,citecolor=blue,linkcolor=blue,urlcolor=blue]{hyperref}
\usepackage{overpic}
\usepackage{array}
\usepackage{nicefrac}
\begin{document}
\title{Electrical tuning of the magnetic properties of 2D  magnets: the case of ${\rm Cr}_2{\rm Ge}_2{\rm Te}_6$}
\author{Guido Menichetti}
\email{guido.menichetti@df.unipi.it}
\affiliation{Dipartimento di Fisica dell'Universit\'a di Pisa, Largo Bruno Pontecorvo 3, I-56127 Pisa,~Italy}
\affiliation{Istituto Italiano di Tecnologia, Graphene Labs, Via Morego 30, I-16163 Genova,~Italy}
\author{Matteo Calandra}
\affiliation{Department of Physics, University of Trento, Via Sommarive 14, 38123 Povo,~Italy}
\affiliation{Istituto Italiano di Tecnologia, Graphene Labs, Via Morego 30, I-16163 Genova,~Italy}
\affiliation{Sorbonne Université, CNRS, Institut des Nanosciences de Paris, UMR7588, F-75252 Paris,~France}
\author{Marco Polini}
\affiliation{Dipartimento di Fisica dell'Universit\'a di Pisa, Largo Bruno Pontecorvo 3, I-56127 Pisa,~Italy}
\affiliation{ICFO-Institut de Ci\`{e}ncies Fot\`{o}niques, The Barcelona Institute of Science and Technology, Av. Carl Friedrich Gauss 3, 08860 Castelldefels (Barcelona),~Spain}

\begin{abstract}
Motivated by growing interest in atomically-thin van der Waals magnetic materials, we present an {\it ab initio} theoretical study of the dependence of their magnetic properties on the electron/hole density $\rho$ induced via the electrical field effect. By focusing on the case of monolayer ${\rm Cr}_2{\rm Ge}_2{\rm Te}_6$ (a prototypical 2D Ising ferromagnet) and employing a hybrid functional, we first study the dependence of the gap and effective mass on the carrier concentration $\rho$. We then investigate the robustness of magnetism by studying the dependencies of the exchange couplings and magneto-crystalline anisotropy energy (MAE) on $\rho$. In agreement with experimental results, we find that magnetism displays a bipolar electrically-tunable character, which is, however, much more robust for hole ($\rho>0$) rather than electron ($\rho<0$) doping. Indeed, the MAE vanishes for an electron density $\rho\approx - 7.5 \times 10^{13}~{\rm e} \times {\rm cm}^{-2}$, signalling the failure of a localized description based on a Heisenberg-type anisotropic spin Hamiltonian. This is in agreement with the rapid increase of the coupling between fourth-neighbor atoms with increasing electron density.
\end{abstract}
\maketitle

\section{Introduction}
\label{sect:intro}

The family of atomically-thin ``beyond graphene'' materials  is now very large~\cite{Geim2013,Mounet2018} and contains also magnetic crystals~\cite{Santos2022,Shabbir2018,Burch2018,Gong2019,Gibertini2019,Soriano2020r,Yang2020,Wei2021}, 
including ${\rm Cr}_2{\rm Ge}_2{\rm Te}_6$~\cite{Gong2017,Wang2018,Kim2019}, ${\rm Cr}{\rm I}_3$~\cite{Huang2017,Wang2018a,Klein2018,Huang2018,Song2018,Jiang2018a,Kim2018,Jin2018,Jiang2018b,Song2019,Thiel2019,Klein2019,Soriano2021,Stavric2023}, ${\rm Fe}_3{\rm Ge}{\rm Te}_2$~\cite{FGT2018_A,FGT2018_B}, ${\rm Cr}{\rm Br}_3$~\cite{CrBr32018,CrBr32019}, and ${\rm Cr}{\rm Cl}_3$~\cite{CrCl32021}. Given their two-dimensional (2D) nature, these materials are being actively investigated as building blocks in van der Waals (vdW) heterostructures to fabricate novel spintronic devices with ultra-small footprint~\cite{Karpiak2019, Zollner2020,Cardoso2018}. In addition, their 2D nature paves the way for achieving an unprecedented control of magnetism via the electrical field effect. 

Though most of these magnets behave as 2D semiconductors, very few groups~\cite{Huang2018,Jiang2018a,Zhang2020,Wang2018,Verzhbitskiy2020,Zhuo2021,Wu2023} have been able to fabricate devices displaying field-effect-transistor-type behavior. Experimental work~\cite{Jiang2018a} on monolayer ${\rm Cr}{\rm I}_3$, for example, revealed that the magnetic properties of this material can be modulated electrically. In fact, hole (electron) doping strengthens (weakens) the magnetic order. Furthermore, an electron doping $\rho \sim -2.5\times 10^{13}~{\rm e} \times {\rm cm}^{-2}$ in bilayer ${\rm Cr}{\rm I}_3$ induces a transition from an antiferromagnetic to a ferromagnetic ground state. This magnetic switching seems to be fully dominated by electrostatic effects and the antiferro-to-ferro phase transition appears to be facilitated by the formation of magnetic polarons~\cite{Soriano2020a}. Similarly, Zhang et al.~\cite{Zhang2020} were able to tune the antiferromagnetic resonances of bilayer ${\rm Cr}{\rm I}_3$ by carrier doping, confirming that electrostatic gating is a powerful tool to modulate various magnetic properties.

Electrostatic gating has also employed to modulate the magnetic properties of ${\rm Cr}_2{\rm Ge}_2{\rm Te}_6$. Few-layer ${\rm Cr}_2{\rm Ge}_2{\rm Te}_6$ exhibits bipolar gate tunability, although much stronger effects have been observed for electron-type doping~\cite{Wang2018,Verzhbitskiy2020,Zhuo2021}. As expected, the $I$-$V$ characteristics of the fabricated field-effect transistors (FETs) indicate a stronger tunability with decreasing the device thickness. In the window of doping $\rho \in [-7,+7] \times 10^{12}~{\rm e} \times {\rm cm}^{-2}$, the Curie critical temperature $T_{\rm C}$ is almost constant, $T_{\rm C}\sim 55~{\rm K}$~\cite{Wang2018}, while for a heavy electron doping of  $\rho \sim -4\times 10^{14}~{\rm e}\times {\rm cm}^{-2}$ a large enhancement of $T_{\rm C}$---up to $T_{\rm C}\sim 200~{\rm K}$---is observed. Heavy electron doping can also switch the sign of the magnetic anisotropy energy (MAE), leading to a change in the magnetic easy axis, from out-of-plane to in-plane. Verzhbitskiy et al.~\cite{ Verzhbitskiy2020} observed that, at high electron densities, the sheet resistance of few-layer ${\rm Cr}_2{\rm Ge}_2{\rm Te}_6$ decreases with increasing temperature, indicating a metallic character. From Hall effect measurements, they inferred the occurrence of an insulator-to-metal transition at $\rho \sim -4\times~10^{14}~{\rm e} \times {\rm cm}^{-2}$. 

On purely theoretical grounds, we conclude by stating that doping a 2D magnetic insulator opens also a wealth of intriguing possibilities in the realm of 2D plasmonics~\cite{Plantey2021}. Indeed,  it has been demonstrated that  spin waves couple to the charge collective modes (i.e.~plasmons) of the itinerant electron system~\cite{Ghosh2022}. These plasmon-magnon interactions naturally arise from the exchange interaction between the itinerant carriers and the localized magnetic moments hosted by the lattice and do not require spin-orbit coupling to exist. Other mechanisms explaining the origin of plasmon-magnon coupling in 2D magnetic crystals have been proposed in the recent literature. In the case of a topological insulator (TI), Efimkin and Kargarian~\cite{Efimkin2021} proposed that the helical nature of spin-momentum locking of Dirac-type surface states is responsible for magnon-plasmon coupling (for earlier work on coupling between charge and spin modes on the surface of a TI see also Ref.~\cite{Raghu2010}). Costa et al.~\cite{Costa2023} and Dyrda et al.~\cite{Dyrdal2023} showed that plasmon-magnon coupling can arise from the interaction between the electromagnetic field of plasmon oscillations and localized spins in a magnet via the direct Zeeman coupling and inverse spin galvanic effect, respectively. The presence of itinerant charge carriers induced via the electrical field effect is the common ingredient in all these works~\cite{Ghosh2022,Efimkin2021,Raghu2010,Costa2023,Dyrdal2023} on plasmon-magnon interactions.

Despite this body of literature, we are unaware of theoretical investigations of the doping dependence of the magnetic properties of 2D vdW materials. There are two ways of studying such dependence. One procedure, dubbed as ``rigid doping'', assumes that the 2D material under study is uniformly doped and overall charge neutrality is compensated through a jellium background. This approach is plagued by an intrinsic problem: no external electric field, which can be greater than $\approx 1.5~{\rm Volt}/$\AA,  is included in the corresponding calculations. As a result, the induced charge density distributes uniformly throughout the simulation cell, even in the vacuum region far from the 2D material and, in the case of thicker flakes, in the bulk of the sample.

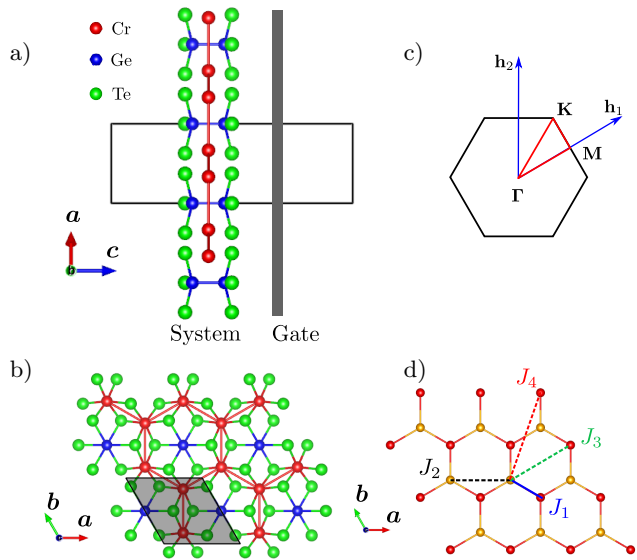
\begin{figure}[t]
\begin{overpic}[width=\columnwidth]{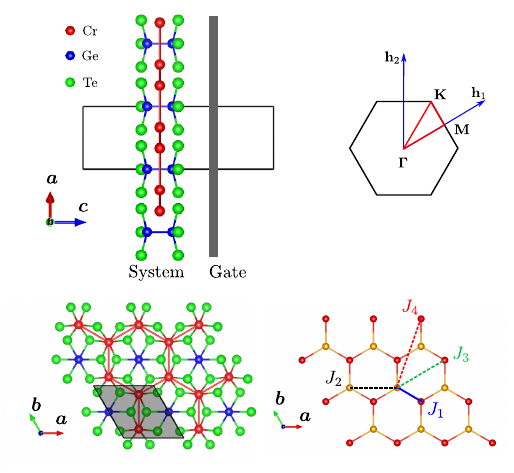} 
\put(1,200){\rm{a)}}
\put(1,80){\rm{b)}}
\put(150,200){\rm{c)}}
\put(150,80){\rm{d)}}
\end{overpic}
\caption{(Color online) a) Monolayer ${\rm Cr}_2{\rm Ge}_2{\rm Te}_6$ placed in front of a metal gate. When a uniform charged plane is placed in front  of the system, the latter is charged with the same amount of opposite charge.  b) Top view of monolayer ${\rm Cr}_2{\rm Ge}_2{\rm Te}_6$. In grey, the rhombohedral unit cell. c) The corresponding Brillouin zone with the high-symmetry points used for the calculation of the electronic band structure~\cite{Grosso2014}.
d) Intra-layer exchange couplings between Cr atoms (orange and red refer to nonequivalent atoms): $J_{1}$, $J_{2}$, $J_3$, and $J_4$ denote the coupling between nearest-neighbor, second-neighbor, third-neighbor, and fourth-neighbor atoms, respectively.
\label{fig:fig1}}
\end{figure}

The second approach~\cite{Brumme2014,Brumme2015,Sohier2017,Baima2018}, which is used in this Article, relies on an effective parallel-plate capacitor model in which one of the plates is a charged layer (here labeled metal gate) and the other ``plate'' is the 2D material of interest---see Fig.~\ref{fig:fig1}a). Given the atomic-scale distance between the gate and the 2D material in Fig.~\ref{fig:fig1}a), this second approach loyally mimics the physisorbed electrical double layer that forms in real devices where ionic liquids are used to achieve large carrier densities such as those ($|\rho| \sim 10^{14}~{\rm e} \times {\rm cm}^{-2}$) explored in this Article. In this second approach, one is able to reproduce the strongly localized voltage drop at the surface of the 2D material and the consequent charging.

In this Article, we present an {\it ab initio}, density functional theory (DFT)~\cite{Giuliani2005,Grosso2014} study of the doping dependence of the electronic, structural, and magnetic properties of monolayer ${\rm Cr}_2{\rm Ge}_2{\rm Te}_6$, which is a prototypical 2D magnetic semiconductor. To this end,  we do not assume ``rigid doping'' but, rather, employ the field effect configuration depicted in Fig.~\ref{fig:fig1}a). We use a {\it hybrid} functional to properly take into account the role of nonlocal electron-electron interaction effects~\cite{Menichetti2019,Wu2019,Lee2020,Ke2021,Menescardi2021}. As discussed at length in our previous work~\cite{Menichetti2019}, such functional transcends the severe limitations that the commonly used ${\rm LDA}+U$/${\rm GGA}+U$ approaches~\cite{anisimov_prb_1991,Cococcioni2005,Campo2010} have displayed in the context of 2D magnetism. In passing, we note that the experimental works~\cite{Wang2018,Verzhbitskiy2020} on ${\rm Cr}_2{\rm Ge}_2{\rm Te}_6$ cited above contain results of DFT calculations. These, however, concentrated only the role of ``rigid doping'' on bulk ${\rm Cr}_2{\rm Ge}_2{\rm Te}_6$ and were carried out using the ${\rm LDA}+U$/${\rm GGA}+U$ approaches, which neglect the crucial  role  of nonlocal electron-electron interactions. 
 
This Article is organized as following. In Sect.~\ref{sect:theory} we present a brief discussion of our theoretical method (Sect.~\ref{sect:technical_details}) and a description of the device geometry (Sect.~\ref{sect:geometry}). In Sect.~\ref{sect:numerical_results} we present a summary of our main numerical results on the electronic structure of monolayer ${\rm Cr}_2{\rm Ge}_2{\rm Te}_6$ in the presence of an electrostatic gate. Results on the doping dependence of the exchange couplings and MAE are instead reported in Sect.~\ref{sect:Heisenberg}. A summary of our main findings and a brief set of conclusions are finally presented in Sect.~\ref{sect:conclusions}. A wealth of additional numerical results is reported in Appendix~\ref{app:extra_numerical_results}.

\section{Theoretical approach}
\label{sect:theory}
In this Section, we first discuss the computational tools we have used (Sect.~\ref{sect:technical_details}) and then (Sect.~\ref{sect:geometry}) present the details of the device geometry (crystal structure of monolayer ${\rm Cr}_2{\rm Ge}_2{\rm Te}_6$ and FET geometry). Uninterested readers can skip Sect.~\ref{sect:technical_details} and jump directly to Sect.~\ref{sect:geometry}.

\subsection{Computational details}
\label{sect:technical_details}

We perform DFT calculations by using the \textsc{Quantum Espresso} (\textsc{QE})~\cite{QE1,QE2,QE3} and \textsc{CRYSTAL17}~\cite{CRYSTAL17,CRYSTAL14} codes, which use plane waves and atom-centered (Gaussian) basis sets, respectively. 
 
For the calculations with~\textsc{QE}  we use the SG15 Optimized Norm-Conserving Vanderbilt (ONCV) pseudopotentials~\cite{Hamann2013,Schlipf2015,Scherpelz2016}, including also the contribution from  spin-orbit coupling (SOC).  We use an energy cutoff up to $70~{\rm Ry}$ for all the calculations.
We adopt the HSE06~\cite{HSE06}, hybrid exchange-correlation (xc) functional, which mixes the Perdew-Burke-Ernzerhof (PBE) exchange~\cite{PBE} with $25\%$ of the exact nonlocal Hartree-Fock exchange, thereby including nonlocal electron-electron interaction effects~\cite{Menichetti2019}. The vdW-D2 correction is included through the method proposed by Grimme~\cite{Grimme2006}. For the Brillouin zone (BZ) integration we employ a Marzari-Vanderbilt smearing~\cite{Marzari1999} 
of $10^{-3}~{\rm Ry}$ with a Monkhorst-Pack (MP)~\cite{MP} ${\bm k}$-point grid with $8\times 8\times 1$ for self-consistent calculations of the charge density.
 
For the calculations with \textsc{CRYSTAL17} we use a 86-411d41 Gaussian all-electron basis set~\cite{Catti1996} with $24$ valence electrons for ${\rm Cr}$ and a double-zeta basis set with an effective core pseudo-potential 
(m-cc-pVDZ-PP)~\cite{Heyd2005} with $24$ valence electrons for ${\rm Te}$ and $22$ valence electrons for ${\rm Ge}$. The charge density integration over the BZ are performed using a uniform  
$24\times 24$  MP ${\bm k}$-point grid. When we use a super-cell, we scale the  MP grid size to assure the same accuracy as in the single-cell calculations. 
  
In order to evaluate fully-relativistic electronic band structures with the HSE06 hybrid xc functional, we used the \textsc{Wannier90}~\cite{W90} code. In fact, at odd with the case of semi-local functionals depending only on the density or the density and its gradient, band structure calculations in the case of hybrid functionals require the knowledge of the electronic wavefunctions in all points where the bands need to be calculated.
  However, at the end of the self-consistent run, the wavefunctions are known only on a uniform grid in the BZ. Thus, an interpolation scheme is needed to extract the electronic structure along high symmetry directions.

The \textsc{Wannier90} code allows us to overcome this problem interpolating the electronic band structure using Maximally-Localized Wannier Functions (MLWFs)~\cite{Marzari1997,Souza2001}  extracted from the DFT calculations with \textsc{QE}. As starting guess for the ``Wannierisation" procedure, we project the Bloch states onto trial localised atomic-like orbitals: $d$-orbitals for the ${\rm Cr}$ atoms, $p$-orbitals for ${\rm Te}$  atoms, and $p_z$-orbital for ${\rm Ge}$ atoms. We made this choice  analyzing the composition of the density of states around the Fermi energy, as shown in Fig.~\ref{fig:fig2}a).

We use \textsc{CRYSTAL17} to evaluate the exchange coupling parameters to reduce the computational cost due to the size of the super-cell and therefore the high number of atoms. In particular, we evaluate the magnetic exchange couplings, with the HSE06 hybrid functional, in a $4\times 4$ super-cell with 160 atoms.  We have been able to switch from \textsc{QE} to \textsc{CRYSTAL17} after testing the consistency of the results with both codes, comparing electronic band structures, relaxed atomic positions, and charge distribution in the FET setup.
 
Finally, we use the \textsc{VESTA}~\cite{VESTA} and \textsc{Xcrysden}~\cite{Crysden} codes to visualize the geometrical structure and BZs---see Fig.~\ref{fig:fig1}. 

\subsection{Geometrical structure and FET Setup}
\label{sect:geometry}
Bulk ${\rm Cr}_2{\rm Ge}_2{\rm Te}_6$ forms a layered structure with monolayers separated by a vdW gap~\cite{Menichetti2019}. Each monolayer---Figs.~\ref{fig:fig1}a) and b)---is formed by edge-sharing ${\rm Cr}{\rm Te}_6$ octahedra where the ${\rm Ge}$ pairs are located in the hollow sites formed by the octahedra honeycomb. In Fig.~\ref{fig:fig1}a) we present the FET device geometry we have studied. It consists of monolayer ${\rm Cr}_2{\rm Ge}_2{\rm Te}_6$ placed in front of a metal gate, which is modeled as a charged plane. The  internal, three-layer structure of monolayer ${\rm Cr}_2{\rm Ge}_2{\rm Te}_6$ therefore acquires the same amount of opposite charge, yielding a finite electric field in the region between the ``system'' and the gate~\cite{Brumme2014,Brumme2015,Baima2018}---see Fig.~\ref{fig:fig1s} in Appendix~\ref{app:extra_numerical_results}. Finally, in Fig.~\ref{fig:fig1}c) we illustrate the BZ of monolayer ${\rm Cr}_2{\rm Ge}_2{\rm Te}_6$.

For the calculation of the electronic and magnetic properties of monolayer ${\rm Cr}_2{\rm Ge}_2{\rm Te}_6$ we use the hexagonal unit cell shown in  Fig.~\ref{fig:fig1}b). We consider a super-cell with about $15$~\AA~of vacuum along the $\hat{\bm z}$-direction between periodic images. We have decided to adopt the experimental lattice parameters to compare on equal footing the results obtained via \textsc{QE} with those obtained via \textsc{CRYSTAL17}~\cite{Menichetti2019}. Using the experimental lattice constants~\cite{Carteaux1995,Siberchicot1996} $a=b=6.8275$~\AA, we relaxed the atomic positions until the maximum force on all atoms was less than $10^{-4}~{\rm Ry}/{\rm Bohr}$. Relaxation of the atomic positions has been carried out in the ferromagnetic phase using using the PBE xc functional. We repeat such
procedure for each doping $\rho$, although we have not observed significant structural variations with changes in this parameter.

Throughout this Article,  $\rho$, measured in units of ${\rm e} \times {\rm cm}^{-2}$, denotes the doping charge density, 
where ${\rm e}=|-e|>0$ is the magnitude of the elementary charge $-e$.  This means that $\rho < 0$ ($\rho > 0$) denotes electron (hole) doping. We explore the following range of doping: $\rho \in [- 1,+1] \times 10^{14}~{\rm e}\times {\rm cm}^{-2}$.

\section{Electronic structure}
\label{sect:numerical_results}
In this Section we present our main results for the electronic structure of  monolayer ${\rm Cr}_2{\rm Ge}_2{\rm Te}_6$ in the FET device geometry. 
\begin{figure*}[tp]
\centering
\begin{tabular}{cc}
  \begin{overpic}[unit=1mm,width=\columnwidth]{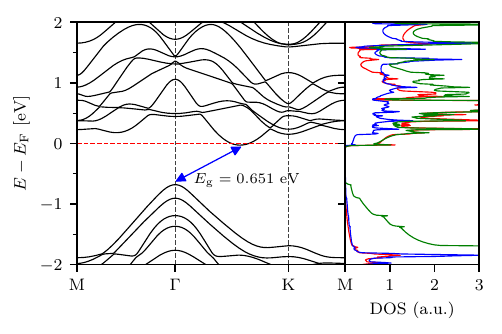} 
\put(3,52){\rm{a)}}
\end{overpic} &
\begin{overpic}[unit=1mm,width=\columnwidth]{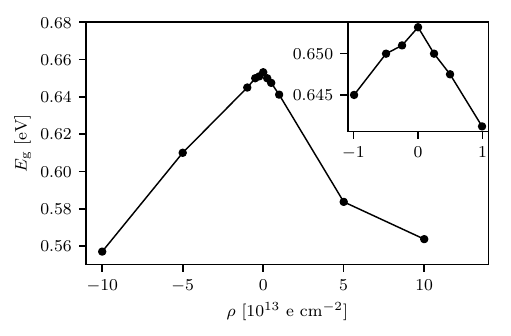}
\put(3,52){\rm{b)}} 
\end{overpic} \\
\begin{overpic}[unit=1mm,width=\columnwidth]{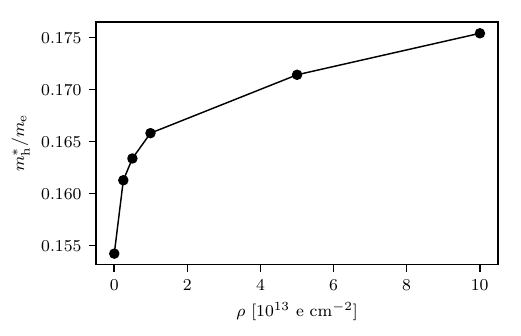} 
\put(3,52){\rm{c)}}
\end{overpic} &
\begin{overpic}[unit=1mm,width=\columnwidth]{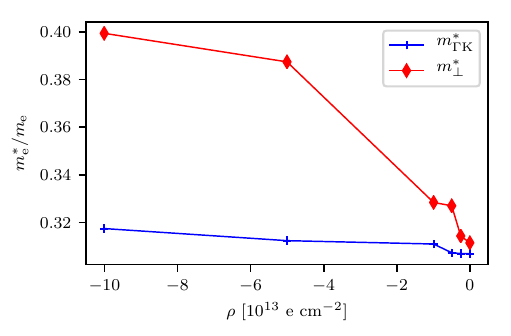}
\put(3,52){\rm{d)}} 
\end{overpic} 
\end{tabular}
 \caption{(Color online) Panel a) Relativistic band structure and projected density of states (DOS) of monolayer ${\rm Cr}_2{\rm Ge}_2{\rm Te}_6$ in the ferromagnetic state. Data in this panel  refer to an electron charge density  $\rho = -2.5\times 10^{12}~{\rm e~ cm}^{-2}$~(corresponding to $-0.01~{\rm e}/{\rm cell}$) and have been obtained with the HSE06 hybrid functional.  In the DOS panel, colors refer to the DOS as projected onto the atomic orbitals of the various atoms (as in Fig.~\ref{fig:fig1}): Cr-$3d$ (red), Ge-$4p$ (blue), and Te-$5p$ (green). The HSE06 hybrid functional yields an insulating ferromagnetic state with an indirect gap $E_{\rm g} \sim 0.651~{\rm eV}$. b) Dependence of the indirect electronic gap $E_{\rm g}$ on the doping charge $\rho$. In the inset a zoom of the results relative to the interval $\rho \in [-1,1] \times 10^{13}$~${\rm e~cm}^{-2}$ is shown. c) Dependence of the effective mass (in unit of the bare electron mass in vacuum $m_{\rm e}$) for holes ($m^*_{\rm h}$) as a function of  $\rho$. d) Same as in panel c) but for electrons: $m^*_{\Gamma{\rm K}}$ ($m^*_{\perp}$) refers to the electron effective mass evaluated at the minimum of the conduction band along the direction $\Gamma{\rm K}$ (orthogonal to $\Gamma{\rm K}$) of the BZ. In panels b)-d), filled symbols represent the calculated points while solid lines are just guides to the eye.
 \label{fig:fig2}}
\end{figure*}

In Fig.~\ref{fig:fig2}a) we show the relativistic electronic band structure of monolayer ${\rm Cr}_2{\rm Ge}_2{\rm Te}_6$ in the ferromagnetic state, with an electron doping of $\rho = - 2.5\times 10^{12}~{\rm e~cm}^{-2}$ $(-0.01~{\rm e}/{\rm cell})$. We find an almost rigid translation of the band profile as compared to the undoped case~\cite{Menichetti2019}. This is in stark contrast with the case of  transition-metal dichalcogenides~\cite{Brumme2015}, where significant differences between ``rigid doping'' and field effect calculations have been reported.

The density of states (DOS), projected onto the atomic orbitals, is also reported in Figs.~\ref{fig:fig2}a). Our results for the DOS demonstrate that the main contribution to the electronic bands near the Fermi energy comes from the ${\rm Cr}-3d$ and ${\rm Te}-5p$ orbitals, which are strongly hybridized~\cite{Menichetti2019}. Such hybridization is very weakly affected by the doping charges. We remind the reader that the ${\rm Te}$ atoms have a fundamental role in stabilizing the ferromagnetic phase of ${\rm Cr}_2{\rm Ge}_2{\rm Te}_6$ because they mediate super-exchange interactions in the ${\rm Cr}$-${\rm Te}$-${\rm Cr}$ bonds, as per the Goodenough-Kanamori rule~\cite{Goodenough1955,Kanamori1959}. This was also confirmed in a recent angle-resolved photoemission spectroscopy study on bulk ${\rm Cr}_2{\rm Ge}_2{\rm Te}_6$~\cite{Watson2020}.

As we will see below in Sect.~\ref{sect:Heisenberg}---and in agreement with experimental results on few-layer ${\rm Cr}_2{\rm Ge}_2{\rm Te}_6$~\cite{Wang2018,Verzhbitskiy2020}---the magnetic properties of monolayer ${\rm Cr}_2{\rm Ge}_2{\rm Te}_6$ change in a much stronger fashion for electron-type (rather than hole-type) doping. Indeed, the magnetic moment localized on the ${\rm Cr}$ atom increases linearly with the doping charge for both dopings, but the rate of change is much lower for hole-type doping---see Fig.~\ref{fig:MagneticMoment} in Appendix~\ref{app:extra_numerical_results}. This behavior can be explained by looking again at the DOS in Fig.~\ref{fig:fig2}a): by adding electrons to the system, states of the ${\rm Cr}$ atom are immediately populated thereby yielding an increase of the its magnetic moment. On the contrary,  a hole-type doping influences mainly the ${\rm Te}$ $5p$ orbitals, leading to a much weaker effect on the ${\rm Cr}$ magnetic moment (and, in turn, to a much weaker tunability of magnetism for holes).

In Fig.~\ref{fig:fig2}b) we plot the dependence of the magnitude of the (indirect) electronic band gap $E_{\rm g}$ on doping $\rho$. We clearly see an approximately {\it linear} decrease of $E_{\rm g}$ with increasing $\rho$, with $E_{\rm g}$ decreasing of about $10~{\rm meV}$ from its undoped ($\rho=0$) value ($\left. E_{\rm g}\right|_{\rho=0} \sim 0.66~{\rm eV}$) to its value at $\rho \pm10^{14}~{\rm e}\times {\rm cm}^{-2}$ (i.e.~$E_{\rm g} \sim 0.56~{\rm eV}$).

Due to our interest in the ${\bm q}$-dispersion of plasmon modes of doped 2D vdW magnets (and their coupling to spin waves)~\cite{Ghosh2022}, we also evaluated  the effective mass of holes (electrons)  via a parabolic fitting of the top (bottom) of the valence (conduction) band. While the hole effective mass $m^*_{\rm h}$ is isotropic near the Fermi level, the electron effective mass $m^*_{\rm e}$ is not. In fact, the Fermi surface near the bottom of the first conduction band has an ellipsoidal shape with the short axis along the $\Gamma{\rm K}$ direction of the BZ. We therefore decompose the electron effective mass $m^*_{\rm e}$ in two components: $m^*_{\Gamma{\rm K}}$ and $m^*_{\perp}$, which are evaluated by fitting the minimum of the conduction band along the $\Gamma{\rm K}$ direction and the direction orthogonal to it, respectively. In Fig.~\ref{fig:fig2}c) (Fig.~\ref{fig:fig2}d)) we report our results for $m^*_{\rm h}$ ($m^*_{\rm e}$). Electrostatic doping increases $m^*_{\rm h}$ and $m^*_{\perp}$, while $m^*_{\Gamma{\rm K}}$ remains approximately constant.

\section{Exchange couplings and magneto-crystalline anisotropy}
\label{sect:Heisenberg}
We now turn to discuss about the dependence on $\rho$ of the magnetic properties of monolayer ${\rm Cr}_2{\rm Ge}_2{\rm Te}_6$. 
\begin{figure*}[t]
\begin{tabular}{cc}
\begin{overpic}[width=\columnwidth]{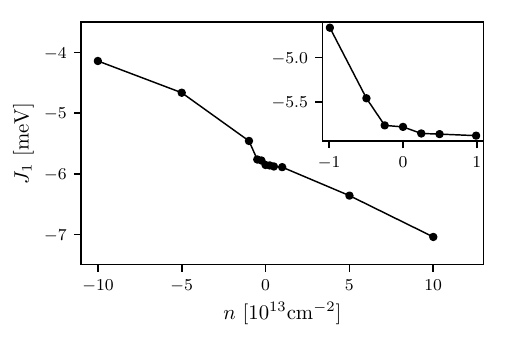}
\put(1,150){\rm{a)}}
\end{overpic} &
\begin{overpic}[width=\columnwidth]{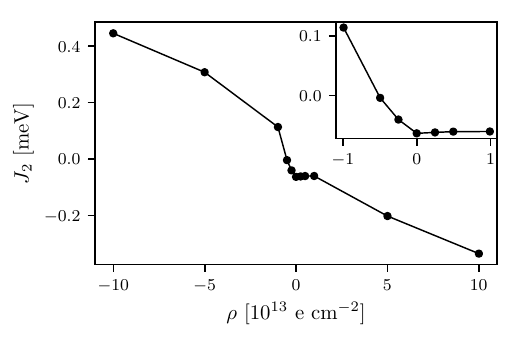}
\put(1,150){\rm{b)}}
\end{overpic} \\
\begin{overpic}[width=\columnwidth]{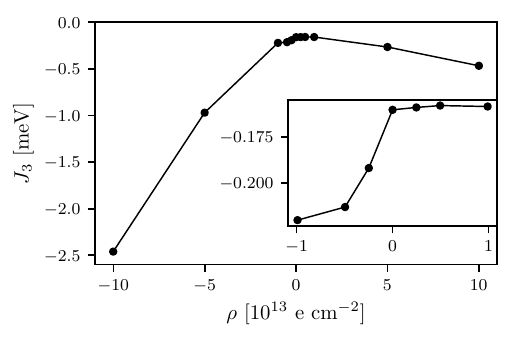}
\put(1,150){\rm{c)}}
\end{overpic} &
\begin{overpic}[width=\columnwidth]{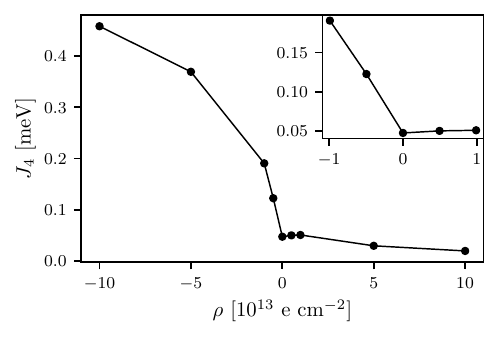}
\put(1,150){\rm{d)}}
\end{overpic} 
\end{tabular}
\caption{Intra-layer exchange coupling parameters $J_i$---defined in Fig.~\ref{fig:fig1}---as functions of the doping charge $\rho$. The inset in every panel shows a zoom of the results relative to the interval $\rho \in [-1,+1] \times 10^{13}$~${\rm e~cm}^{-2}$. In all panels, filled circles are the calculated data points while solid lines are just guides to the eye. \label{fig:fig3}}
\end{figure*}

In order to understand the microscopic properties of magnetic materials it is crucial to map the {\it ab-initio} results onto an effective spin model. In the case of localized magnetic moments, the Heisenberg spin Hamiltonian can be used. In our previous work~\cite{Menichetti2019}, we calculated all the relevant exchange couplings (and the MAE)  for monolayer, bilayer, and bulk ${\rm Cr}_2{\rm Ge}_2{\rm Te}_6$, with the HSE06 hybrid functional. Here, we greatly extend this previous study by calculating how the intra-layer exchange couplings of monolayer ${\rm Cr}_2{\rm Ge}_2{\rm Te}_6$ change with $\rho$. On the one hand, these calculations allow us to understand the range of values of $\rho$ where a localized description of magnetism based on the Heisenberg spin Hamiltonian is valid. On the other hand, our results allow us to understand the mechanism that enables electrical control of  magnetism. Indeed, a high level of doping may affect the super-exchange mechanism that stabilizes the ferromagnetic ground state.

The spin Hamiltonian we use has the following form (see Ref.~\cite{Menichetti2019} and references therein to earlier work)
\begin{equation}\label{eq:Heisenberg+MAE}
\hat{\cal H}_{\rm spin}=\sum_{i<j} J_{ij}\hat{\bm S}_i\cdot \hat{\bm S}_j+\sum_{i}A_i\hat{S}^2_{iz}~.
\end{equation}
\begin{itemize}
\item[i)] The first term on the right-hand side of Eq.~(\ref{eq:Heisenberg+MAE}) is the rotationally-invariant Heisenberg Hamiltonian with exchange couplings $J_{ij}$, the sum over $i, j$ running over all the ${\rm Cr}$ pairs without double counting. 
\item[ii)] The second term is the MAE, the sum over $i$ running over all the ${\rm Cr}$ sites. (We remind the reader that this term is crucial to stabilize long-range magnetic order in 2D~\cite{Mermin1966,Torelli2018} since it lifts the rotational invariance of the first term in Eq.~(\ref{eq:Heisenberg+MAE}), effectively allowing to bypass the Mermin-Wagner theorem~\cite{Mermin1966}.) The MAE, which stems from SOC~\cite{Bruno1989,Li2014}, can be calculated~\cite{Gong2017,Li2018,Li2014,Menichetti2019} by looking at the total energy difference---obtained through self-consistent calculations in the presence of SOC---between the configuration with all spins aligned along the $\hat{\bm z}$ direction (i.e.~perpendicular to the plane hosting monolayer ${\rm Cr}_2{\rm Ge}_2{\rm Te}_6$) and that with all spins aligned along the $\hat{\bm x}$ or $\hat{\bm y}$ direction (i.e.~parallel to the plane hosting monolayer ${\rm Cr}_2{\rm Ge}_2{\rm Te}_6$).  A positive (negative) sign of the MAE means that the system is an easy-plane (easy-axis) ferromagnet. 
\item[iii)] We emphasize that Eq.~(\ref{eq:Heisenberg+MAE}) does not contain a Dzyaloshinski-Moriya (DM) interaction term of the form $\sum_{i<j}{\bm D}_{ij}\cdot(\hat{\bm S}_i\times\hat{\bm S}_j)$. This is certainly fully justified in the case of a free-standing monolayer ${\rm Cr}_2{\rm Ge}_2{\rm Te}_6$ due to the presence of an inversion symmetry center~\cite{Moriya1960}. However, the presence of an electric field orthogonal to the atomically-thin system breaks such symmetry and a non-zero DM should be, at least in principle, taken into account~\cite{Vishkayi2020}. However, we do expect such contribution to be very small in magnitude, on the order of $\sim {\rm \mu eV}$, as reported in Ref.~\onlinecite{Vishkayi2020} for the case of ${\rm Cr}{\rm I}_3$, because, as discussed in Sect.~\ref{sect:numerical_results}, the electrical field effect geometry in the present case induces an almost rigid shift of the band structure and no significant structural changes.
\end{itemize} 

Microscopic calculations of the exchange couplings $J_{ij}$ are more subtle than the MAE. As discussed at length in Ref.~\cite{Menichetti2019}, an accurate method to calculate $J_{ij}$ for small-gap semiconductors and metals is the four-state mapping analysis (FSMA) approach~\cite{Whangbo2003,Xiang2011,Xiang2013a,Sabani2020}. The FSMA approach considers one specific magnetic pair at a time in a super-cell. Such method is accurate when the super-cell used for the calculations is large enough. In fact, we can monitor the accuracy of the calculated exchange coupling by increasing the super-cell size~\cite{Menichetti2019}. In the present case of a doped system, we observed that we need at least a $3\times 3$ super-cell of monolayer ${\rm Cr}_2{\rm Ge}_2{\rm Te}_6$ to have converged results with doping charge in the range $\rho \in [-1,+1]~\times 10^{13}$~${\rm e~cm}^{-2}$. For values of $\rho$ outside this interval we were forced to use a $\sqrt{13}\times \sqrt{13}$ super-cell---see Fig.~\ref{fig:convergence}-\ref{fig:convergence2} in Appendix~\ref{app:extra_numerical_results}.

\subsection{Exchange couplings}

In Fig.~\ref{fig:fig3} we present our results for the intra-layer exchange couplings $J_{1}, \dots, J_4$ (defined in Fig.~\ref{fig:fig1}), which are plotted as functions of the doping charge $\rho$. These results have been calculated with the HSE06 hybrid functional, but without the inclusion of SOC. The reason is that we do not expect that SOC contributes appreciably to the exchange couplings as the magnetic moments of the ${\rm Cr}$ atoms do not chance significantly with the inclusion of SOC~\cite{Menichetti2019}.  For $\rho \in [-1,+1]~\times 10^{13}$~${\rm e~cm}^{-2}$, we observe much stronger effects for electron-type doping. For this type of doping, indeed, all the exchange couplings $J_i$ display an almost linear dependence on $\rho<0$: see the insets in Figs.~\ref{fig:fig3}a)-d). For hole-type doping $J_i$ are roughly independent of $\rho$. These results are in agreement with the DOS analysis discussed earlier in Sect.~\ref{sect:numerical_results} and the experimental results reported in Ref.~\onlinecite{Wang2018}.
\begin{figure}[t]
\begin{overpic}[width=\columnwidth]{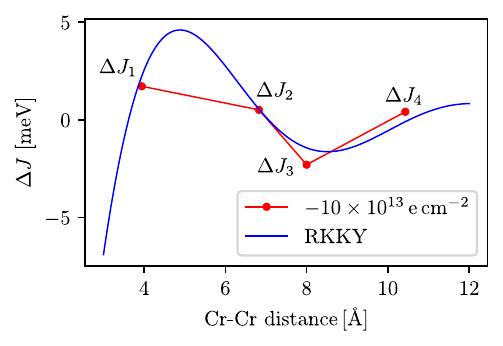}
\end{overpic} 
\caption{(Color online) The difference $\Delta J$ (filled red circles) between the exchange couplings evaluated at finite $\rho$ and the same quantities evaluated for $\rho=0$ (i.e.~in the undoped system). On the horizontal axis we have displayed the ${\rm Cr} - {\rm Cr}$ distance. Results in this plot refer to $\rho = - 10^{14}$~${\rm e~cm}^{-2}$. The oscillating behaviour of $\Delta J$ is fitted (blue curve) with an RKKY interaction for a 2D electron gas model~\cite{Giuliani2005,Litvinov1998,Zhang2020_PRB}.    
\label{fig:fig4}}
\end{figure}

The situation becomes more complicated for larger values of $\rho$: see main panels in Figs.~\ref{fig:fig3}a)-d). Increasing $\rho<0$ to values larger than $-10^{13}$~${\rm e~cm}^{-2}$, $J_1$ is significantly reduced, reaching  $J_1 \approx -4~{\rm meV}$ for $\rho=-1\times 10^{14}$~${\rm e~cm}^{-2}$. In the same interval, $J_2$ changes sign, while remaining much smaller than $J_1$ in magnitude, thereby not altering the ferromagnetic order. 
For large values of $\rho<0$, $J_3$ becomes large and negative (thereby contributing to the stabilization of ferromagnetism), reaching $J_3 \approx -2~{\rm meV}$ for $\rho=-1\times 10^{14}$~${\rm e~cm}^{-2}$. Such a large value of $J_3$ suggests a change in character of magnetism, from a strongly localized to a rather delocalized type, for heavy electron doping. This is the reason why we evaluated also the fourth-neighbor exchange coupling  $J_4$~\cite{CONVERGENCE}. This coupling is on the order of $0.01~{\rm meV}$ in the undoped system (and for small values of $\rho$) but it becomes significantly large for large values of $\rho<0$. We find $J_4 \approx -0.6~{\rm meV}$  for $\rho= -10^{14}$~${\rm e~cm}^{-2}$.  

These trends of the exchange couplings suggest that, for large values of $\rho<0$, a long-range, carrier-mediated interaction emerges (on top of the local, super-exchange mechanism). Examples include interactions mediated by the double-exchange mechanism~\cite{Verzhbitskiy2020,Zener1951} or Ruderman-Kittel-Kasuya-Yosida (RKKY) interactions~\cite{Grosso2014,Giuliani2005,Aristov1997,Litvinov1998}. These phenomena arise due to the delocalized nature of the conduction bands, which stems from the hybridization of the ${\rm Cr}-3d$ orbitals with the ${\rm Te}-5p$ orbitals~\cite{Menichetti2019}. In Fig.~\ref{fig:fig4} we plot the difference $\Delta J$ between the exchange couplings in the presence and absence of doping charge, as a function of the ${\rm Cr}-{\rm Cr}$ bond length. An RKKY model for 2D systems~\cite{Giuliani2005,Aristov1997,Litvinov1998,footnoteRKKY,Zhang2020_PRB,Tang2021} fits well our numerical data. Such qualitative validity of the RKKY fit corroborates the metallic character of magnetism observed experimentally~\cite{Verzhbitskiy2020} for large values of $\rho<0$. From a theoretical point of view, much more work is however needed to precisely establish which long-range scenario better describes the magnetic properties of  monolayer ${\rm Cr}_2{\rm Ge}_2{\rm Te}_6$ for large values of $\rho<0$.

The dependence of the exchange couplings on $\rho$ for holes ($\rho>0$) is not as complicated as for electrons. As clear in Fig.~\ref{fig:fig3}, after a small plateau for small values of $\rho>0$, the exchange couplings increase with increasing $\rho>0$, leading to a more stable ferromagnetic order. The largest changes with $\rho>0$ are observed in $J_1$ and  $J_2$, while $J_{3}$ and $J_{4}$ do not change significantly, even for large values of $\rho>0$. This suggests that, contrary to what discussed above for electrons, magnetic order retains its localized character with increasing $\rho>0$, even at high hole densities---see Figs.~\ref{fig:fig5s}-\ref{fig:fig6s} in Appendix~\ref{app:extra_numerical_results} to see the stability of ferromagnetic phase in monolayer ${\rm Cr}_2{\rm Ge}_2{\rm Te}_6$ with respect to changes in  $\rho$ and results for the exchange couplings obtained in the ``rigid doping'' approach.

\begin{figure}[t]
\centering
\begin{tabular}{c}
\begin{overpic}[unit=1mm,width=\columnwidth]{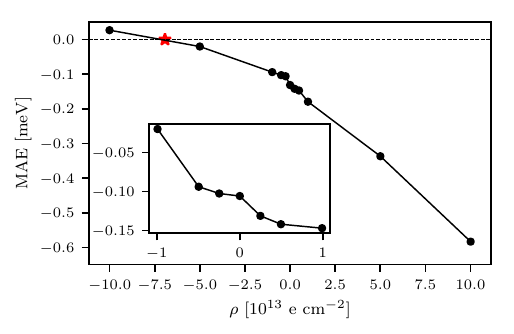} 
\put(3,50){\rm{a)}}
\end{overpic} \\
\begin{overpic}[unit=1mm,width=\columnwidth]{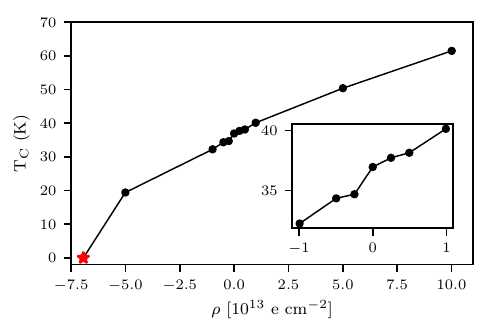}
\put(3,50){\rm{b)}} 
\end{overpic}
\end{tabular} 
 \caption{(Color online)  Panel a) MAE (per Cr atom) as a function of the doping charge $\rho$. The red star denotes the value of electron doping $\rho_{\rm c}$ at which the MAE vanishes (i.e.~$\rho_{\rm c} \approx -6.4 \times 10^{13}~{\rm e~cm}^{-2}$). At this value of doping the Curie temperature $T_{\rm C}$ vanishes.  Panel b) Curie temperature $T_{\rm C}$ as a function of doping charge $\rho$. In the inset we report a zoom of the results relative to the interval $\rho \in [-1,1] \times 10^{13}~{\rm e~cm}^{-2}$. In both panels, filled circles are the calculated points---according to Eqs.~(\ref{eq:TC1})-(\ref{eq:TC2})---while solid lines are just guides to the eye.
 \label{fig:fig5}}
\end{figure}
\subsection{Magneto-crystalline anisotropy energy}

In Fig.~\ref{fig:fig5}a) we presents our results for the dependence of the MAE (per ${\rm Cr}$ atom) on $\rho$. We clearly see that the MAE increases in absolute value for hole-type doping ($\rho>0$). Once again, the situation for electron-type doping is dramatically different. With increasing $\rho<0$, the MAE first decreases and then changes sign at $\rho_{\rm c} \sim -6.4 \times 10^{13}$~${\rm cm}^{-2}$. (This critical value $\rho_{\rm c}$ has been obtained by interpolating in a linear fashion between two results for the MAE, one obtained at $\rho=-10^{14}~{\rm e~cm}^{-2}$ and one at $\rho=-5 \times 10^{13}~{\rm e~cm}^{-2}$.)  This signals a change in the magnetic easy axis, from out-of-plane to in-plane. At this value of electron doping, long-range magnetic order ceases to exist due to the Mermin-Wagner theorem~\cite{Mermin1966}.  We also estimate the Curie temperature $T_{\rm C}$ by considering only the largest exchange coupling (i.e.~$J_1$) and the MAE. We use an expression for $T_{\rm C}$ proposed by the authors of Ref.~\cite{Torelli2018}, which reads as following:
\begin{equation}\label{eq:TC1}
T_{\rm C}=T^{\rm Ising}_{\rm C}f\left(\dfrac{A}{J_1}\right)
\end{equation}
with
\begin{equation}\label{eq:TC2}
f(x)=\tanh^{1/4}\left[\dfrac{6}{N_{\rm nn}}\log (1+\gamma x) \right]~.
\end{equation}
Here, $A$ coincides with the MAE, $N_{\rm nn}$ is the number of nearest neighbors ($3$ in our case), $\gamma = 0.033$, and $T^{\rm Ising}_{\rm C}= 1.52 S^2 J_1 / k_{\rm B}$.

Using the calculated dependencies of $J_1$ and ${A}$ on $\rho$ inside Eq.~\eqref{eq:TC1} we obtain the Curie temperature trend reported in Fig.~\ref{fig:fig5}b), which is in excellent qualitative agreement with the experimental results of Refs.~\onlinecite{Wang2018,Verzhbitskiy2020} for thin films. Indeed, $T_{\rm C}$ is almost constant for small doping, i.e.~for $\rho \in [-1,+1]~\times 10^{13}~{\rm e~cm}^{-2}$ while for heavy hole doping $T_{\rm C}$ is almost twice as large as the critical temperature for the undoped case. A more quantitative analysis between our calculations and the above mentioned experimental results is beyond the scope of the present work for two reasons. On the one hand, we focus on monolayer ${\rm Cr}_2{\rm Ge}_2{\rm Te}_6$ while the results of Refs.~\onlinecite{Wang2018,Verzhbitskiy2020} refer to thin films ($T_{\rm C}$ has a strong dependence on the thickness of the sample~\cite{Menichetti2019,Gong2017}). On the other hand, a truly quantitative comparison would require the inclusion of $J_2$ in a theory of $T_{\rm C}$, while formula (\ref{eq:TC1}) includes only the MAE and $J_1$. 

\section{Summary and conclusions}
\label{sect:conclusions}
In summary, in this Article we have reported on an extensive  density functional theory study of the electronic and magnetic properties of the atomically-thin magnetic material ${\rm Cr}_2{\rm Ge}_2{\rm Te}_6$, in the presence of an electrical field effect. 

Taking into account nonlocal eletron-electron interactions through the use of the HSE06 hybrid functional~\cite{Menichetti2019}, we have calculated the relativistic electronic band structure of monolayer ${\rm Cr}_2{\rm Ge}_2{\rm Te}_6$ showing its rigidity under the application of an external electric field induced by a nearby metal gate. We have calculated the dependence on the doping charge $\rho$ of the electronic band gap---see Fig.~\ref{fig:fig2}b)---and hole/electron effective masses---see Fig.~\ref{fig:fig2}c)-d). We have then computed the dependence of the magnetic properties on $\rho$, i.e.~the exchange couplings up to the fourth-neighbour one (see Fig.~\ref{fig:fig3}), the magneto-crystalline anisotropy energy (see Fig.~\ref{fig:fig5}a)), and the approximate Curie temperature $T_{\rm C}$ (see Fig.~\ref{fig:fig5}b)). The obtained results have been compared with experimental data, when available.

The non-negligible values of the fourth-neighbour exchange coupling $J_4$ at high electron doping signal a change in the character of magnetism and raise questions on the underlaying microscopic mechanism. Qualitatively, our data for heavy electron dopings can be described by an RKKY-type interaction---see Fig.~\ref{fig:fig4}. The authors of Ref.~\cite{Verzhbitskiy2020}, however, suggest that, at high doping, a carrier-mediated double-exchange mechanism kicks in. Understanding on theory grounds the best microscopic mechanism responsible for magnetism at high electron dopings is beyond the scope of the present Article and is left for future work. More in general, much more experimental work on few-layer ${\rm Cr}_2{\rm Ge}_2{\rm Te}_6$ is needed to assess the microscopic validity of nonlocal functionals for atomically-thin 2D materials.

\acknowledgments

This work was supported by the European Union's Horizon 2020 research and innovation programme under grant agreement No.~881603 - GrapheneCore3, under the Marie Sklodowska-Curie grant agreement No.~873028 - Hydrotronics.  and by the MUR - Italian Minister of University and Research under the ``Research projects of relevant national interest  - PRIN 2020''  - Project No.~2020JLZ52N, title ``Light-matter interactions and the collective behavior of quantum 2D materials (q-LIMA)''.
M. C. acknowledges financial support from the Italian ``Centro Nazionale di Ricerca in High Performance Computing, Big Data and Quantum Computing'', supported by the EU-funds of Italy’s Recovery and Resilience Plan. Finally, we acknowledge  the ``IT center'' of the University of Pisa, the HPC center of the Fondazione Istituto Italiano di Tecnologia (IIT, Genova), and the allocation of computer resources from CINECA, through the ``ISCRA C'' projects ``HP10C9JF51'', ``HP10CI1LTC'', ``HP10CY46PW'', and ``HP10CMQ8ZK". 
\clearpage 

\setcounter{section}{0}
\setcounter{equation}{0}%
\setcounter{figure}{0}%
\setcounter{table}{0}%

\setcounter{page}{1}

\renewcommand{\thetable}{A\arabic{table}}
\renewcommand{\theequation}{A\arabic{equation}}
\renewcommand{\thefigure}{A\arabic{figure}}
\renewcommand{\bibnumfmt}[1]{[A#1]}
\renewcommand{\citenumfont}[1]{A#1}

\appendix
\onecolumngrid

\section{Additional numerical results}
\label{app:extra_numerical_results}
In this Appendix we present additional numerical results.

\begin{figure}[h]
\begin{tabular}{cc}
\begin{overpic}[width=0.45\columnwidth]{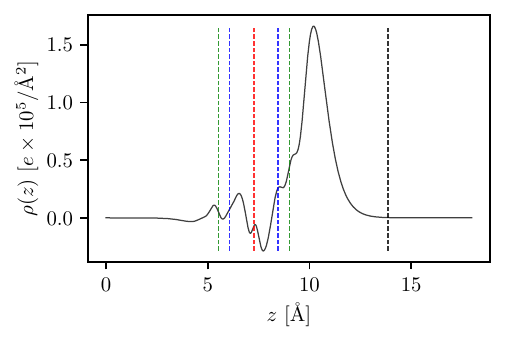}
\put(0,150){\rm{a)}}
\end{overpic} & 
\begin{overpic}[width=0.45\columnwidth]{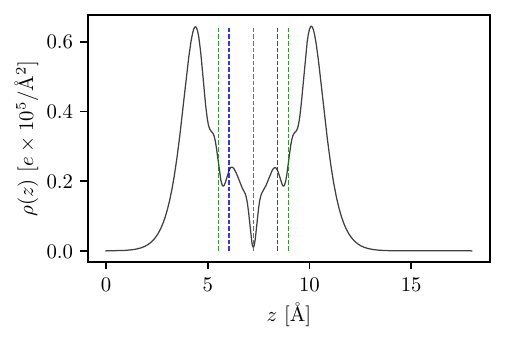}
\put(0,150){\rm{b)}}
\end{overpic}
\end{tabular}
\caption{(Color online) Doping density $\rho(z)$ of monolayer ${\rm Cr}_2{\rm Ge}_2{\rm Te}_6$ as a function of $z$. The vertical dashed lines indicate the position of the atoms, the color coding being identical to that reported in Fig.~\ref{fig:fig1}a) of the main text. In panel a) we clearly see that the doping charge mainly accumulates in the inner layers of ${\rm Cr}_2{\rm Ge}_2{\rm Te}_6$ closer to the gate. In panel b) we plot the same quantity as in panel a) but for the case of ``rigid doping''.
\label{fig:fig1s}}
\end{figure}

In Fig.~\ref{fig:fig1s}a) we plot  the doping density $\rho(z)$ due to the presence of the metal gate, after integrating the three-dimensional doping density over the 2D system's area. Corresponding results for the case of rigid doping are reported in Fig.~\ref{fig:fig1s}b).

\begin{figure}[h]
\begin{overpic}[width=0.4\columnwidth]{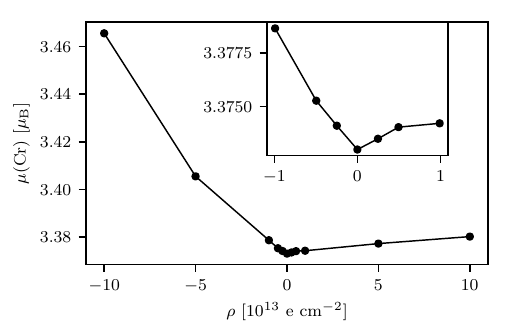} 
\end{overpic} 
\caption{Calculated magnetic moment of ${\rm Cr}$ (in units of the Bohr's magneton $\mu_{\rm B}$) as a function of the doping charge $\rho$. In the inset we report a zoom of the results relative to the interval $\rho \in [-1,1] \times 10^{13}~{\rm e~cm}^{-2}$.
\label{fig:MagneticMoment}}
\end{figure}

In Fig.~\ref{fig:MagneticMoment} we report the calculated magnetic moment of ${\rm Cr}$ as a function of the doping charge density $\rho$. 

\begin{figure}[h]
\begin{tabular}{ccc}
\begin{overpic}[width=0.32\columnwidth]{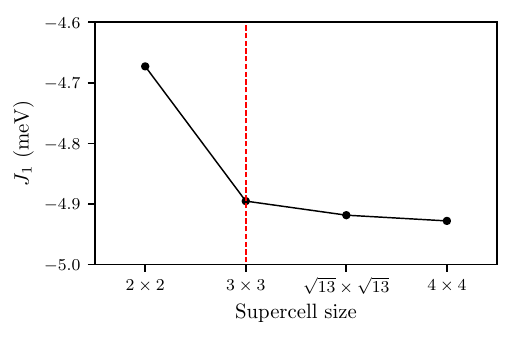}
\put(0,100){\rm{a)}}
\end{overpic} &
\begin{overpic}[width=0.32\columnwidth]{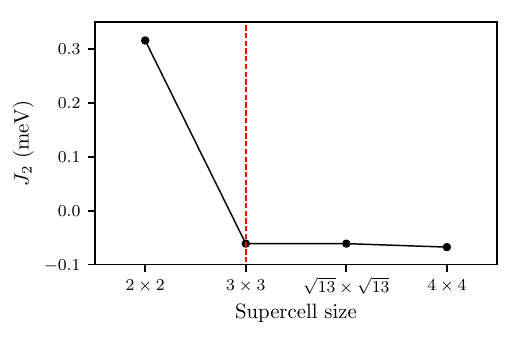}
\put(0,100){\rm{b)}}
\end{overpic} &
\begin{overpic}[width=0.32\columnwidth]{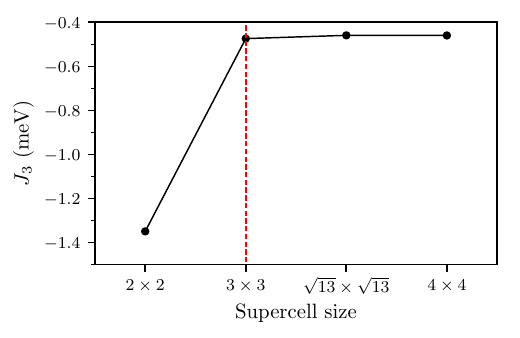}
\put(0,100){\rm{c)}} 
\end{overpic} 
\end{tabular}
\caption{(Color online) Convergence of the intra-layer exchange coupling parameters $J_1$, $J_2$, and $J_3$ of monolayer ${\rm Cr}_2{\rm Ge}_2{\rm Te}_6$ with the size of the super-cell. Results in this figure refer to a doping $\rho \approx -5\times 10^{12}~{\rm cm}^{-2}$ and have been obtained in the framework of the PBE functional and in the electrical field effect configuration depicted in Fig.~\ref{fig:fig1}a) of the main text. The red vertical dashed lines denote the values of the super-cell size used in this Article. 
\label{fig:convergence}}
\end{figure}

\begin{figure}[h]
\begin{overpic}[width=0.4\columnwidth]{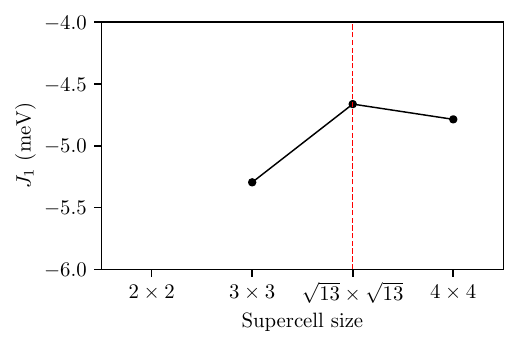} 
\end{overpic} 
\caption{(Color online) Convergence of the intra-layer exchange coupling parameter $J_1$ with the size of the super-cell. Results in this figure refer to a doping $\rho \approx -5\times 10^{13}~{\rm cm}^{-2}$ (i.e.~ten times higher than in Fig.~\ref{fig:convergence}) and have been obtained in the framework of the HSE06 hybrid xc functional and in the electrical field effect configuration depicted in Fig.~\ref{fig:fig1}a) of the main text. The red vertical dashed line denotes the values of the super-cell size used in this Article.
\label{fig:convergence2}}
\end{figure}

In Fig.~\ref{fig:convergence} and~\ref{fig:convergence2} we illustrate a summary of our convergence tests in the context of the calculations of the intra-layer exchange coupling parameters $J_i$. All the necessary details are reported in the corresponding figure captions.

\begin{figure}[h]
\begin{overpic}[width=0.4\columnwidth]{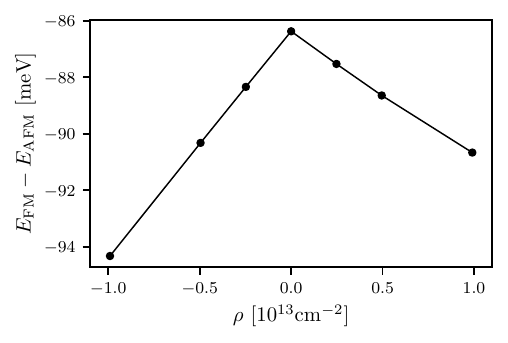} 
\end{overpic} 
\caption{The difference $E_{\rm FM} - E_{\rm AFM}$ between the total energy of the ferromagnetic ($E_{\rm FM}$) and anti-ferromagnetic ($E_{\rm AFM}$) phases is plotted as a function of $\rho$. \label{fig:fig5s}}
\end{figure}

In Fig.~\ref{fig:fig5s} we present an analysis of the stability of ferromagnetic phase in monolayer ${\rm Cr}_2{\rm Ge}_2{\rm Te}_6$ with respect to changes in  $\rho$.

\begin{figure}[h]
\begin{tabular}{ccc}
\begin{overpic}[width=0.33\columnwidth]{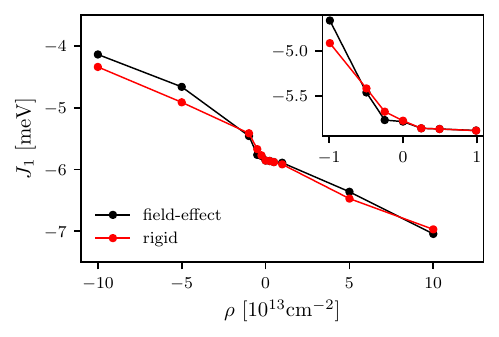}
\put(1,100){\rm{a)}}
\end{overpic} &
\begin{overpic}[width=0.33\columnwidth]{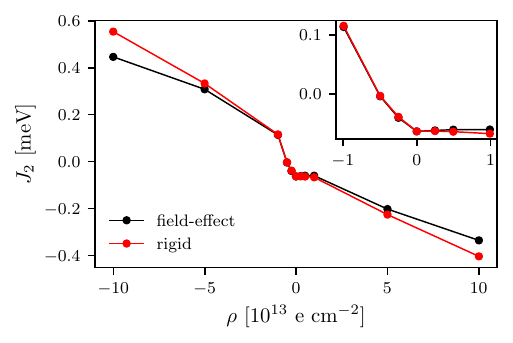}
\put(1,100){\rm{b)}}
\end{overpic} &
\begin{overpic}[width=0.33\columnwidth]{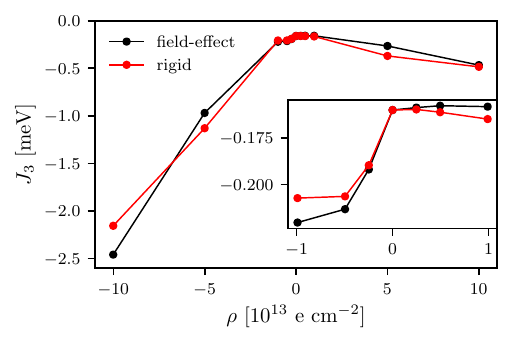}
\put(1,100){\rm{c)}}
\end{overpic}
\end{tabular}
\caption{(Color online) The three intra-layer exchange coupling parameters $J_1$, $J_2$, and $J_3$ are plotted as functions of doping $\rho$. Results in this plot have been obtained with the HSE06 hybrid functional. The inset in each panel contains results in the interval $\rho \in [-1,1] \times 10^{13}~{\rm e~cm}^{-2}$. Results obtained in the simplified  ``rigid'' doping'' approach (red symbols) are compared to the ones obtained in the electrical field effect approach (black symbols).  
\label{fig:fig6s}}
\end{figure}

In Fig.~\ref{fig:fig6s} we compare results for $J_1$, $J_2$ and $J_3$ obtained with two different methods: i) black symbols denote results obtained in the electrical field effect geometry sketched in Fig.~\ref{fig:fig1}a) of the main text; ii) red symbols denote results obtained in the ``rigid doping'' approach.


\begin{thebibliography}{70}
\bibitem{CONVERGENCE}
We also have reached a converged value of $J_4$ in a $\sqrt{13}\times \sqrt{13}$ super-cell.
\end{thebibliography}
\end{document}